%% file: master_2021.tex
\documentclass[tightenlines,superscriptaddress,11pt,notitlepage]{revtex4-2}

\usepackage{amsmath}
\usepackage{amssymb}
\usepackage{amstext}
\usepackage{mathtools}
\usepackage{pifont}
\usepackage{graphicx}
\usepackage{outline}
\usepackage{fancyhdr}
\usepackage{lineno}
\usepackage{enumitem}
\usepackage[margin=1.0in]{geometry}
\usepackage{url}
\usepackage{wrapfig}
\usepackage{titlesec}
\usepackage{fancyhdr}
\usepackage{color}
\usepackage[position=bottom, caption=false]{subfig}

\titlespacing\section{0pt}{12pt plus 4pt minus 2pt}{0pt plus 2pt minus 2pt}
\titlespacing\subsection{0pt}{12pt plus 4pt minus 2pt}{0pt plus 2pt minus 2pt}

\renewcommand*{\paragraph}[2]{\textbf{#2}}

\makeatletter
\renewcommand{\fnum@table}{\textbf{\tablename~\thetable}}
\renewcommand{\fnum@figure}{\textbf{\figurename~\thefigure}}
\makeatother

\makeatletter
\AtBeginDocument{\let\LS@rot\@undefined}
\makeatother

\newcommand\lsim{\mathrel{\rlap{\lower4pt\hbox{\hskip1pt$\sim$}}
    \raise1pt\hbox{$<$}}}
\newcommand\gsim{\mathrel{\rlap{\lower4pt\hbox{\hskip1pt$\sim$}}
    \raise1pt\hbox{$>$}}}

\newcommand{\beq}{\begin{equation}}
\newcommand{\eeq}{\end{equation}}
\newcommand{\bea}{\begin{eqnarray}}
\newcommand{\eea}{\end{eqnarray}}
\newcommand{\bem}{\begin{pmatrix}}
\newcommand{\eem}{\end{pmatrix}}

\begin{document}

\title{\scshape\Large \bf Hyper-Kamiokande Experiment: A Snowmass White Paper\\
\vspace{5mm}
\normalsize Contributed Paper to Snowmass 2021
\vspace{5mm}
%\vskip -10pt
}
\author{J.~Bian}\affiliation{Department of Physics and Astronomy, University of California Irvine, Irvine CA 92697-4575, USA}
\author{F.~Di Lodovico}\affiliation{King's College London, Department of Physics, Strand Building, Strand, London, United Kingdom}
\author{S.~Horiuchi}\affiliation{Center for Neutrino Physics, Virginia Tech, Blacksburg, Virginia 24061, USA}
\author{J.~G.~Learned}\affiliation{Department of Physics and Astronomy, University of Hawaii, Honolulu, Hawaii 96822, USA}
\author{C.~Mariani}\email{Corresponding author: mariani@vt.edu}\affiliation{Center for Neutrino Physics, Virginia Tech, Blacksburg, Virginia 24061, USA}
\author{J.~Maricic}\affiliation{Department of Physics and Astronomy, University of Hawaii, Honolulu, Hawaii 96822, USA}
\author{J.~Pedro~Ochoa~Ricoux}\affiliation{Department of Physics and Astronomy, University of California Irvine, Irvine CA 92697-4575, USA}
\author{C.~Rott}\affiliation{Department of Physics and Astronomy, University of Utah, Salt Lake City, UT 84112, USA}\affiliation{Department of Physics, Sungkyunkwan University, Suwon 16419, Korea}
\author{M.~Shiozawa}\affiliation{University of Tokyo, Institute for Cosmic Ray Research, Kamioka Observatory, Kamioka, Japan}\affiliation{University of Tokyo, Kavli Institute for the Physics and Mathematics of the Universe (WPI), University of Tokyo Institutes for Advanced Study, Kashiwa, Japan}\affiliation{University of Tokyo, Next-generation Neutrino Science Organization, Kamioka, Japan}
\author{M.~B.~Smy}\affiliation{Department of Physics and Astronomy, University of California Irvine, Irvine CA 92697-4575, USA}
\author{H.~W.~Sobel}\affiliation{Department of Physics and Astronomy, University of California Irvine, Irvine CA 92697-4575, USA}
\author{R.~B.~Vogelaar}\affiliation{Center for Neutrino Physics, Virginia Tech, Blacksburg, Virginia 24061, USA}

\collaboration{on behalf of the Hyper-Kamiokande Collaboration}    
\date{\today}

\begin{abstract}
\vspace{2cm}
Hyper-Kamiokande (HK) is the next generation underground water Cherenkov detector that builds on the highly successful Super-Kamiokande (SK) experiment. The 260,000-ton detector has an 8.4 times larger fiducial volume than its predecessor. HK's low energy threshold combined with the very large fiducial volume make the detector unique; HK is expected to acquire an unprecedented exposure of 3.8 Mton-year over a period of 20 years of operation. It has an extremely diverse science program including long-baseline neutrino oscillation measurements, nucleon decay searches, atmospheric neutrinos, neutrinos from the sun and supernova explosions, and neutrinos from other astrophysical origins. Like DUNE, the flagship project of the U.S. high-energy physics program, HK measures fundamental properties of neutrinos such as the search for leptonic CP violation and neutrino physics beyond the Standard Model.
\end{abstract}

\maketitle

\renewcommand{\familydefault}{\sfdefault}
\renewcommand{\thepage}{\roman{page}}
\setcounter{page}{0}

\pagestyle{plain} 
\clearpage
\textsf{\tableofcontents}

\renewcommand{\thepage}{\arabic{page}}
\setcounter{page}{1}

\pagestyle{fancy}

% Set how header/footers look
%\newcommand{\chaptermark}[1]{%
%\markboth{Chapter \thechapter:\# 1}{}}
%\renewcommand{\chaptermark}[1]{%
%\markboth{Chapter \thechapter:\ #1}{}}
\fancyhead{}
%\fancyhead[RO,LE]{\textsf{\footnotesize \thepage}}
%\fancyhead[RO]{\textsf{\footnotesize \thepage}}
%\fancyhead[LO,RE]{\textsf{\footnotesize \rightmark}}
%\fancyhead[LO]{\textsf{\footnotesize \rightmark}}

\fancyfoot{}
%\fancyfoot[RO]{\textsf{\footnotesize Snowmass 2021,\thepage}}
\fancyfoot[RO]{\textsf{\footnotesize \thepage}}
\fancyfoot[LO]{\textsf{\footnotesize Hyper-Kamiokande: A Snowmass White Paper}}
\fancypagestyle{plain}{}

%upper and bottom footer lines
%\renewcommand{\headrule}{\vspace{-4mm}\color[gray]{0.5}{\rule{\headwidth}{0.5pt}}}
\renewcommand{\footrulewidth}{0.4pt}

\newpage
%%%%%%%%%%%%%%%%%%%%%%%%%%%%%%%%%%%%%%%%%%%%%%%%%%%%%%%%%%%%%%
% Narrative
%%%%%%%%%%%%%%%%%%%%%%%%%%%%%%%%%%%%%%%%%%%%%%%%%%%%%%%%%%%%%%
% a brief overview of the HyperK project as a whole (4-6 pages)

\input{overview.tex}

\newpage

\section*{\bf \large References}
\renewcommand{\refname}{References}

\bibliography{overview,oscillation}
\bibliographystyle{apsrev}

%%%%%%%%%%%%%%%%%%%%%%%%%%%%%%%%%%%%%%%%%%%%%%%%%%%%%%%%%%%%%%
\end{document}

%% file: overview.tex
\section{\bf \large Overview}

The HK collaboration is presently formed amongst groups from 19 countries including the United States, whose community has a long history of making significant contributions to the neutrino physics program in Japan. US physicists have played leading roles in the Kamiokande, KamLAND, SK, EGADS, K2K, and T2K programs. Currently, the HK design is being finalized while excavation has already been started. HK data taking is expected to start in 2027.

The US groups are interested in exploiting the complementarity of the HK and DUNE experiments. This complementarity is similar to the one of the T2K and the NO$\nu$A experiments: different baselines, beam energies (narrow-band vs. wide-band), detector technology, detection mechanism and detector size. While both DUNE and HK are also observatories for supernova neutrinos, they are complementary in their detection channels (electron neutrino detection vs. anti-electron neutrino detection).  There is also complementarity in the nucleon decay channels.

 The motivation for pursuing "natural" neutrino sources in addition to measuring neutrino beams remains strong: not only did the study of these neutrino sources lead to many groundbreaking discoveries in the past (including the discovery of neutrino flavor oscillation and mass, measurements of the mixing angles, discovery of the oscillation pattern, etc.), but future measurements of these sources will significantly contribute to determine the mass ordering and support and complement CP violation and other beam-related neutrino measurements.
 
{\vskip 0.25in}
In the following sections we give details on particular physics topics and objectives, highlighting the complementary between the HK and DUNE physics programs:

\begin{enumerate}
\item Nucleon Decay and Dark Matter
\item Solar Neutrinos
\item Supernova Neutrinos, Multi-messenger science, Relic Supernova Neutrinos
\item Atmospheric Neutrinos, Neutrino Mass Ordering, Non-standard interactions and CP violation
\item Accelerator Neutrinos - The J-PARC to Hyper-K long baseline experiment
\end{enumerate}

%%%%%%%%%%%%%%
{\vskip 0.25in}
\noindent
\section{\bf \large Nucleon Decay and Dark Matter}
{\vskip 0.075in}

\begin{figure}[tb!]
    \centering
    \includegraphics[width=1.0\columnwidth]{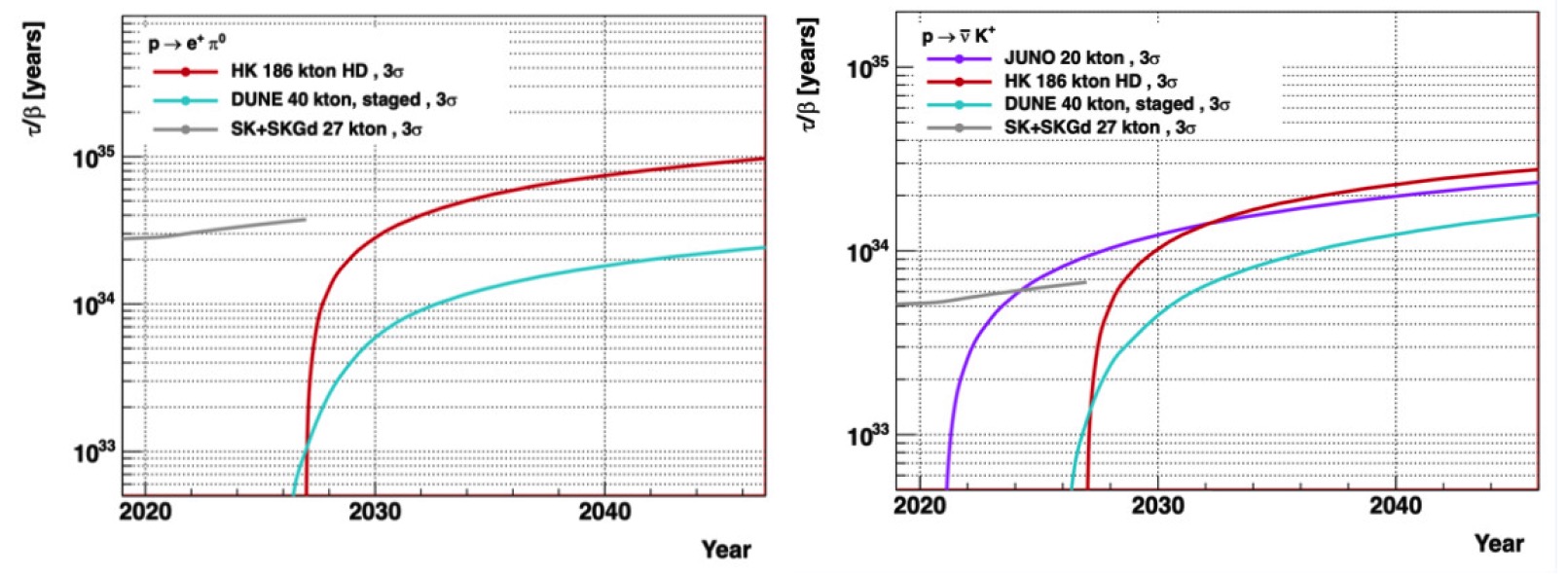}
    \caption{Nucleon decay physics reach of Hyper-K~\cite{Itow:2021rnc}.}
    \label{fig:nucleon_decay}
\end{figure}

HK will be able to provide some of the most stringent tests of the Standard Model. The possibility that baryon number violation holds the key to a number of questions in fundamental physics and cosmology makes the continued search for proton decay and related processes a high priority. We expect the HK experiment to pick up where the SK program leaves off. For example, nucleon decay sensitivities will be extended by one order of magnitude [Fig.~\ref{fig:nucleon_decay}] beyond the current limits and could reveal grand unified theories (GUTs)~\cite{HyperKamiokande:2018ofw}. 

%\begin{figure}[tb!]
%    \centering
%    \includegraphics[width=1.0\columnwidth]{figures/nucleon_decay.jpg}
%    \caption{Nucleon decay physics reach of Hyper-K~\cite{HyperKamiokande:2018ofw}}
%    \label{fig:nucleon_decay}
%\end{figure}

With 20~years of data, Hyper-K will reach a proton decay sensitivity of $10^{35}$~years for $p \rightarrow \pi^0 e^+$ and $3\times 10^{34}$~years for $p \rightarrow \bar{\nu} K^+$. These decay modes are highly complementary to those accessible by liquid argon based detectors: HK has a huge total mass, and lower mass per nucleus (including free protons), but offers less detail in event reconstruction and some particles (e.g. Kaons) are near or below the Cherenkov threshold. HK proton decay searches are similarly complementary to JUNO which is also sensitive to Kaons {\it and} has a low mass per nucleus, but much lower total mass than HK and much better reconstruction capabilities than liquid Argon detectors.

\vspace{1cm}

The search for physics beyond the Standard Model of particle physics is one of the priorities of HK. Following SK's success, the indirect search for dark matter from the Sun, Earth, Galactic center, and halo is expected to continue to provide the most sensitive tests for dark matter nucleon scattering for masses of a few GeV in a model independent way~\cite{Choi:2013eda,SuperKamiokande:2015xms,Danninger:2014xza}. HK will search for boosted dark matter~\cite{SuperKamiokande:2017dch}, asymmetric dark matter~\cite{Murase:2016nwx}, or test models with predominantly hadronic annihilation channels that remain hidden to other neutrino detectors~\cite{Rott:2012qb,Bernal:2012qh,Rott:2015nma}. Searches for dark matter from the Galactic halo will provide some of the best sensitivities to dark matter for annihilation channels with large neutrino yields. 

%%%%%%%%%%%%%%
{\vskip 0.25in}
\noindent
\section{\bf \large Solar Neutrinos}
{\vskip 0.075in}

\par Hyper-K will detect solar neutrino interactions with unprecedented statistical power. Like Super-K, the interaction mode is elastic scattering of $^8$B neutrinos. The analysis of the recoil electron spectrum reveals the underlying neutrino spectrum. In particular, the ``upturn'', the vacuum--MSW transition,  will be detectable at $5\sigma$ ($3\sigma$) based on 10 years of data with an energy threshold of $3.5$~MeV ($4.5$~MeV)~\cite{HyperKamiokande:2018ofw}. In addition to constraining solar neutrino oscillation parameters in the standard oscillation scenario, the  recoil  electron  spectrum provides sensitivity  to  non-standard  neutrino interactions. Hyper-K will also measure the diurnal variation of the solar neutrino interaction rate due to Earth matter effects. In addition to measuring oscillation parameters, this effect is sensitive to non-standard interactions as well. Also the comparison of solar neutrino oscillation parameters with reactor anti-neutrino oscillation parameters tests CPT invariance.
Hyper-K is particularly well suited to measure short-period flux variations in solar neutrinos, realizing a real-time monitoring of the Solar core temperature. Moreover, Hyper-K could achieve the first measurement of hep solar neutrinos, giving new insights in solar physics. Hyper-K solar neutrino measurements would be complementary to possible DUNE measurements using solar $^8$B neutrino absorption on argon: while the neutrino absorption process has better energy correlation between the detected electron and the incident neutrino, elastic scattering avoids the high nuclear energy threshold of $\approx5$ MeV. The cosmogenic background in DUNE will be lower (due to the greater depth), but elastic scattering measurements  have better signal/noise ratio due to the good directional correlation of electrons and neutrinos. 

Low energy interactions (less than $\sim$100~MeV where most electrons will start electromagnetic showers) are more challenging than higher energy events since the light yield in a Cherenkov detector (number of detected photo-electrons per MeV above Cherenkov threshold) is much smaller than in a liquid scintillator (or the ionization yield in a TPC): O(10) detected photo electrons per MeV~\cite{HyperKamiokande:2018ofw}. In fact, for Super-K the light yield is about six per MeV, while for a detector like JUNO it will be around 1300 per MeV~\cite{2022103927}. This makes event reconstruction and radioactive background discrimination more challenging, and results in a higher threshold  compared to JUNO's expected $\sim$2~MeV. However, Hyper-K's strengths are its unmatched size and the ability to reconstruct the charged particle's direction, which makes it more resilient against backgrounds and is an important feature for astrophysics.
 
%%%%%%%%%%%%%%
{\vskip 0.25in}
\noindent
\section{\bf \large Supernova Neutrinos, Multi-messenger Science}
{\vskip 0.075in}

Through the observation of $\sim$10~MeV neutrinos with time, energy and directional information, Hyper-K will take a unique role as multi-messenger observatory. Hyper-K will expand on the successful multi-messenger science program of Super-K~\cite{SuperKamiokande:2016jsv}. It has the potential to detect thermal neutrinos from nearby ($<10$~Mpc) neutron star merger events in coincidence with gravitational waves.

Given sufficient running time, the observation of a signal in HK due to a core collapse supernova is assured~\cite{Hirata:1987hu,Ikeda:2007sa,Nakamura:2016kkl}. It is estimated that the chance of a Type II supernova in the Milky Way galaxy is a few percent per year. What we might learn from such an event is relatively unknown, although there exist many models that make baseline predictions for comparison once we have the data. What is certain is that the statistical signal will dwarf the 19 events observed by IMB and Kamiokande from SN1987A. For a supernova at 10~kpc, somewhere between 16,000 to 64,000 neutrino events above 7~MeV would be detected by Hyper-K in a few seconds, depending on the supernova model. Hyper-Kamiokande will provide the largest sample of the most detailed events, including elastic scatter events that will reveal the direction of the supernova and enable the astronomical community to engage in multi-messenger discoveries. Hyper-K's reach extends to the Andromeda Galaxy M31 ($\sim$780~kpc) with about $\sim$10 events expected per supernova. 

The direction of a supernova at 10~kpc can be reconstructed with an accuracy of about 1$^o$ assuming similar event reconstruction performance as Super-K~\cite{HyperKamiokande:2018ofw}, making Hyper-K essential for distributing early alerts and multi-messenger observations. Further, even a few neutrinos from nearby extra-galactic supernovae can reveal the nature of transients whose mechanism is uncertain~\cite{SuperKamiokande:2000kzn,Dragowsky:2001ax}. 
With tens of thousands of events from a supernova in the Milky Way galaxy, it is anticipated that studying the time, energy, and flavor structure of the burst may reveal not only astrophysical secrets of the core collapse and subsequent explosion, but also information on neutrino properties. The observed burst signal may reveal the effects of neutrino-neutrino interactions in a regime inaccessible by any other means, as well as effects due to beyond-the-Standard-Model physics. Since the matter effect resonance is directly connected to the neutrino mass hierarchy, it is possible that structure in the time-energy spectrum may reveal whether the hierarchy is inverted or normal. If the supernova occurs when both Hyper-K and DUNE are running, the large sample of inverse-beta decay events from Hyper-K and the large sample of electron neutrino charge-current events from DUNE will yield highly complementary flavor information of the core-collapse supernova neutrinos, opening a novel flavor window to study a variety of effects from supernova shock effects to neutrino properties such as the neutrino mass ordering.

{\vskip 0.25in}
\noindent
\section{\bf \large Supernova Relic Neutrinos}
{\vskip 0.075in}
 
\begin{figure}[htb!]
    \centering
    \includegraphics[width=0.7\columnwidth]{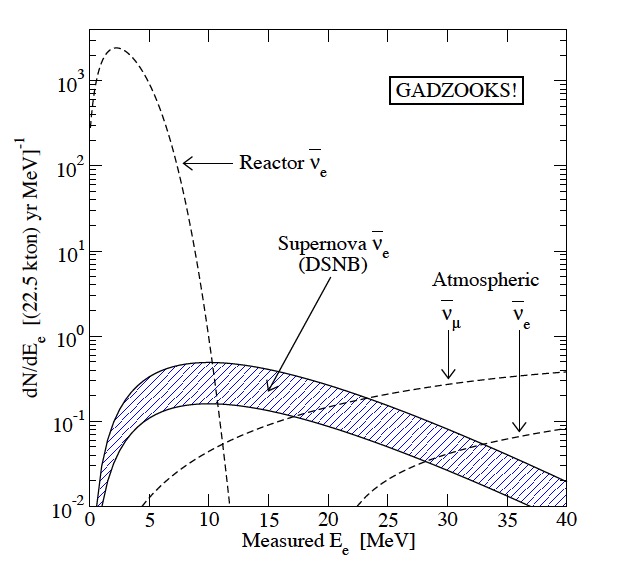}
    \caption{The spectrum of anti-neutrinos due to relic supernova neutrinos compared with backgrounds from atmospheric neutrinos and reactors. The width of the SRN band represents the range of model predictions. The dashed atmospheric neutrino lines reflect the expected background without neutron capture on gadolinium~\cite{HyperKamiokande:2018ofw}.}
    \label{fig:supernovae}
\end{figure}

In addition to being prepared for a gravitational collapse supernova event, we have another way to study supernova neutrinos. Neutrinos emitted by supernovae at a distance beyond Hyper-Kamiokande's ability to uniquely localize in time, comprise the so-called ``diffuse supernova neutrino background'' (DSNB), also known as the ``supernova relic neutrinos'' (SRN). These are one of the few remaining neutrino signals that has not yet been detected by any experiment. When observed, the measured rate of SRN could provide information about stellar collapse, nucleosynthesis, and the rate of star formation in the universe. The addition of Gd to Hyper-Kamiokande or direct comparison between pure water HK and Gd loaded SK would address the backgrounds that have limited prior searches~\cite{Super-Kamiokande:2011lwo}, by allowing the signal selection to require the distinctive signature of inverse beta decay followed by neutron capture. The search window would lie between the background due to nuclear reactor antineutrinos and backgrounds due to atmospheric neutrinos [Fig.~\ref{fig:supernovae}].

%%%%%%%%%%%%%%%%%%%%%%%%%%%%%
{\vskip 0.25in}
\noindent
\section{\bf \large Atmospheric Neutrinos - Neutrino Mass Ordering and Non-standard Interactions}
{\vskip 0.075in}

Precise oscillation measurements with atmospheric neutrinos will be used to study the neutrino mass ordering as well as push limits for Lorentz symmetry violation~\cite{SuperKamiokande:2014exs,IceCube:2017qyp}, non-standard interaction~\cite{SuperKamiokande:2011dam,IceCube:2017zcu}, quantum decoherence~\cite{Stuttard:2020qfv}, sterile neutrino oscillation~\cite{T2K:2019efw,IceCube:2020tka}, and various dark sector particle searches such as heavy neutral lepton~\cite{T2K:2019jwa,Coloma:2019htx}, long lived particle~\cite{Arguelles:2019ziu}, and millicharged particles~\cite{Plestid:2020kdm}.

\begin{figure}[tb!]
\subfloat[]{\includegraphics[width=0.52\columnwidth]{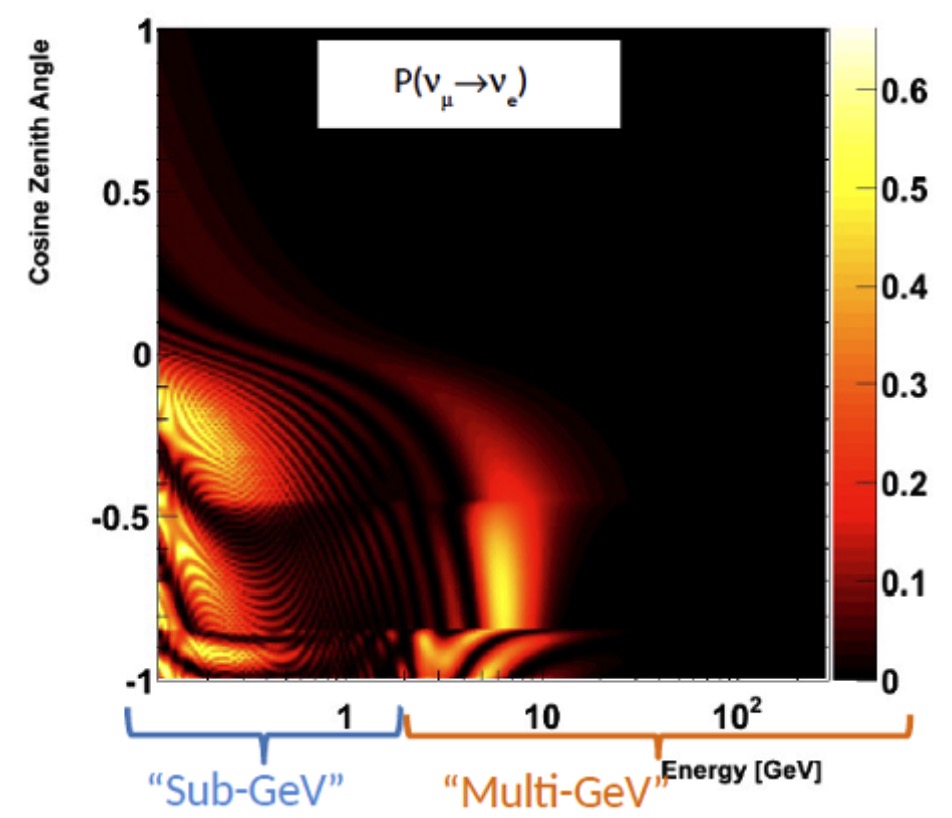}}
\subfloat[]{\includegraphics[width=0.45\columnwidth]{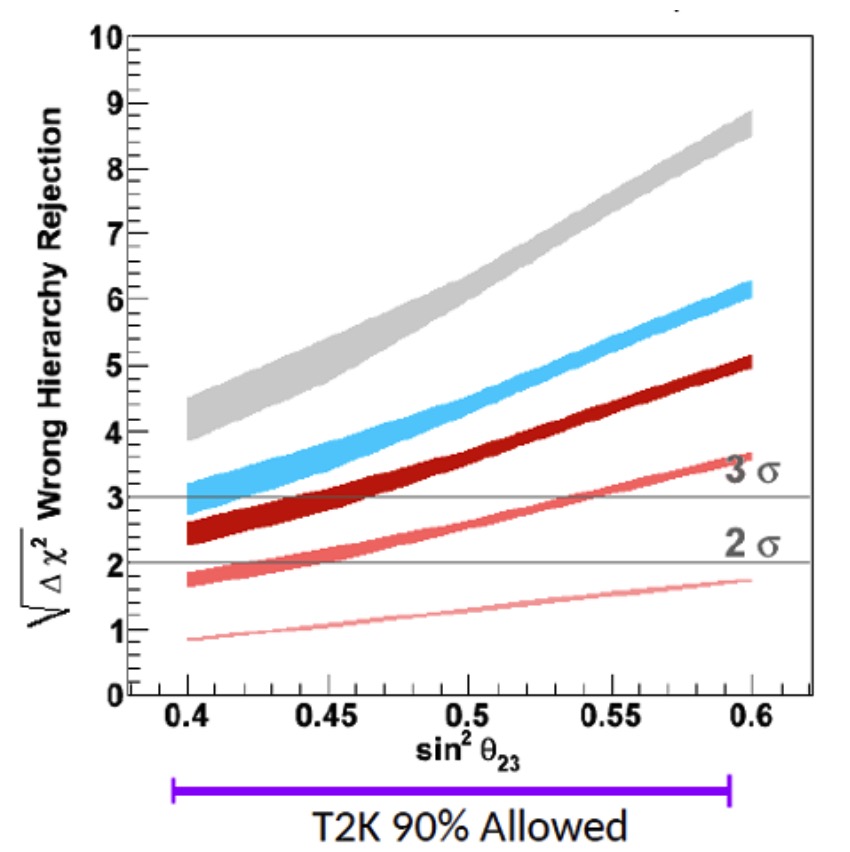}}
\caption{(a) The oscillogram showing oscillation probability from $\nu_\mu$ to $\nu_e$ as a function of energy and cosine zenith angle. (b) The resultant sensitivity for wrong hierarchy rejection as a function of $\sin^2\theta_{23}$ and running time.  Running time is for 1, 5, 10, 15, 30 years. Figures from Ref.~\cite{HyperKamiokande:2018ofw}.}
\label{fig:atmospheric}
\end{figure}

Atmospheric neutrinos span a wide range of energy and path lengths, and are born in both the neutrino and antineutrino form of electron and muon flavor. Neutrinos that travel through the earth pass through a substantial matter density that modifies the vacuum oscillation probability in a predictable way. This feature gives HK the potential to determine the neutrino mass ordering from atmospheric neutrinos alone. Hierarchy sensitivity predominantly comes from the ``Multi-GeV'' upward-going electron neutrinos as seen in Fig.~\ref{fig:atmospheric}. There are resonant oscillations between 2-10~GeV for $\nu_\mu$ or $\bar{\nu_\mu}$ depending on the mass hierarchy.

\begin{figure}[tb!]
\subfloat[]{\includegraphics[width=0.48\columnwidth]{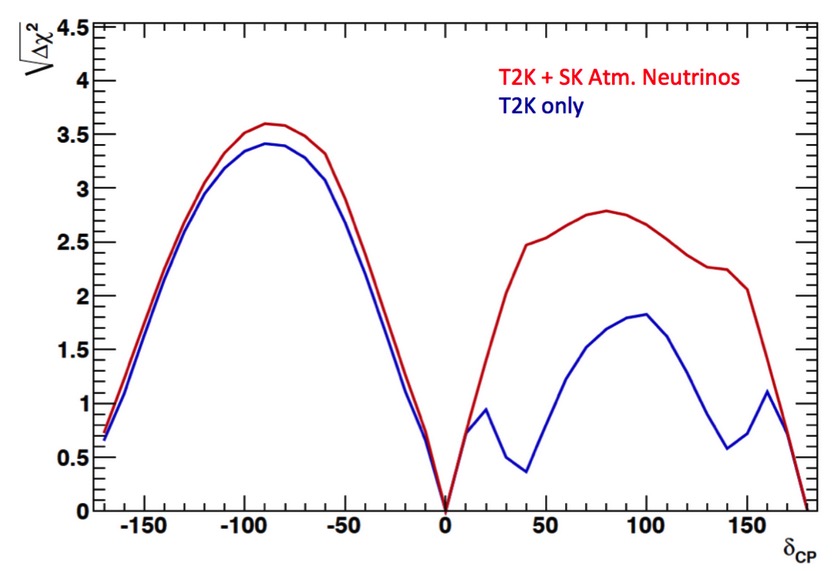}}
\subfloat[]{\includegraphics[width=0.52\columnwidth]{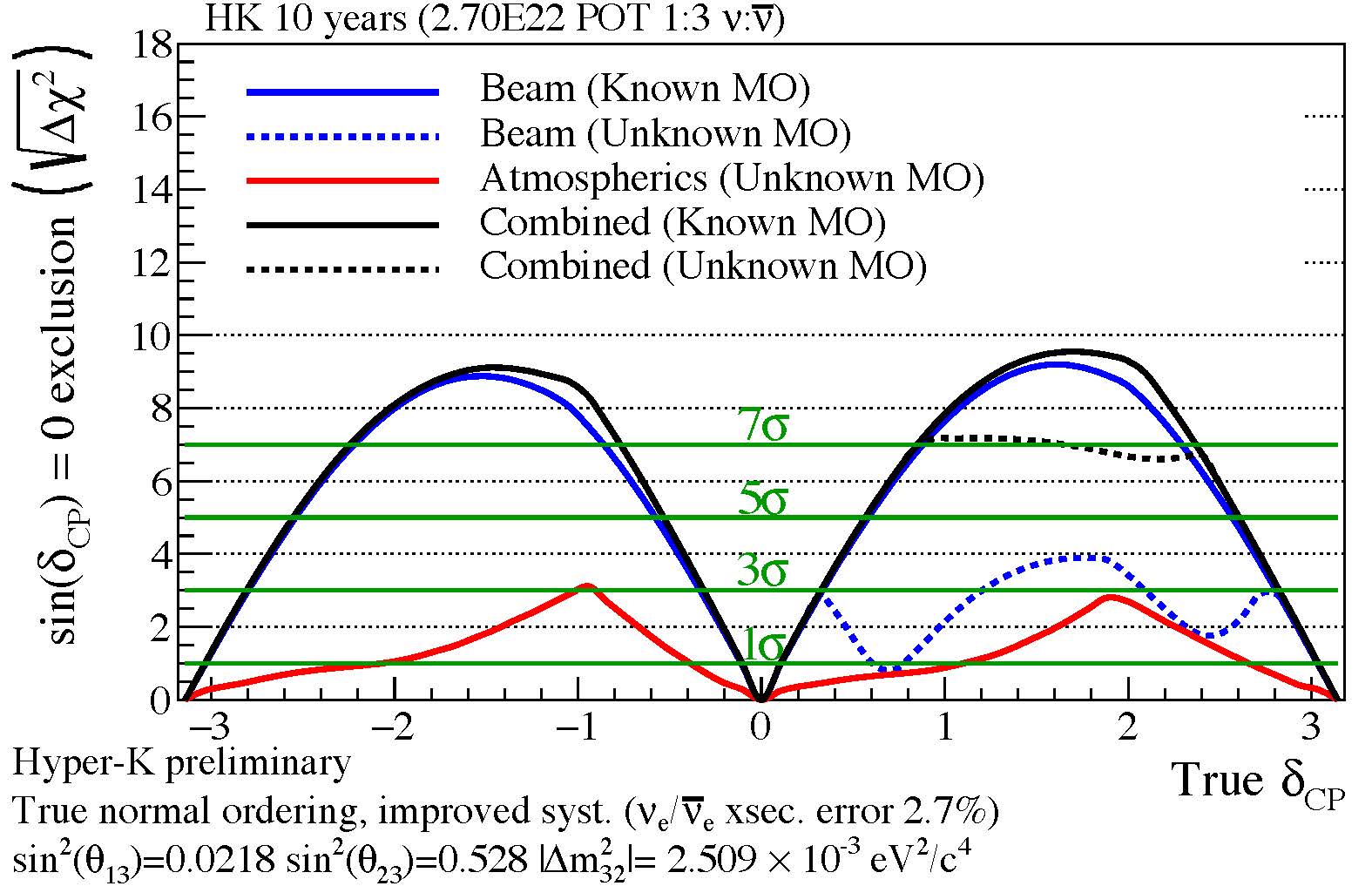}}
\caption{(a) The sensitivity for T2K (solid blue) and T2K$+$SK (solid red) to reject the sin($\delta_{CP}$)=0 hypothesis assuming unknown hierarchy.  (b) Projected significance for $\sin \delta_{CP}$=0 exclusion for Hyper-K assuming normal ordering.}
\label{fig:sensitivity}
\end{figure}

Although atmospheric neutrinos have limited sensitivity to CP-violation relative to the beam measurement, the sensitivity is largely complementary and the addition of atmospheric neutrino data to the beam measurement can improve the $\delta_{CP}$ measurement, particularly in regions of limited beam sensitivity due to degeneracies between oscillation parameters (i.e. mass hierarchy and CP effects). As an example, the sensitivity for T2K compared to T2K $+$ SK is shown in [Fig.~\ref{fig:sensitivity}]. The addition of SK atmospheric neutrino data improves rejection for $0 < \delta_{CP} < \pi$ since the T2K baseline~\cite{Ankowski:2015jya,Ankowski:2016bji}, like the HK baseline, is relatively short and feels no matter-effect.  The final projected sensitivity for HK is shown in Fig.~\ref{fig:sensitivity}.

The unitarity of the PMNS matrix can be tested through $\nu_\tau$ appearance~\cite{Ankowski:2016bji,Benhar:2015wva,Ankowski:2016jdd}. The current determinations of PMNS mixing elements agree between atmospheric neutrinos and long-baseline or reactor experiments, but tensions could appear that reveal new research directions.

\begin{figure}[htb!]
    \centering
    \includegraphics[width=0.75\columnwidth]{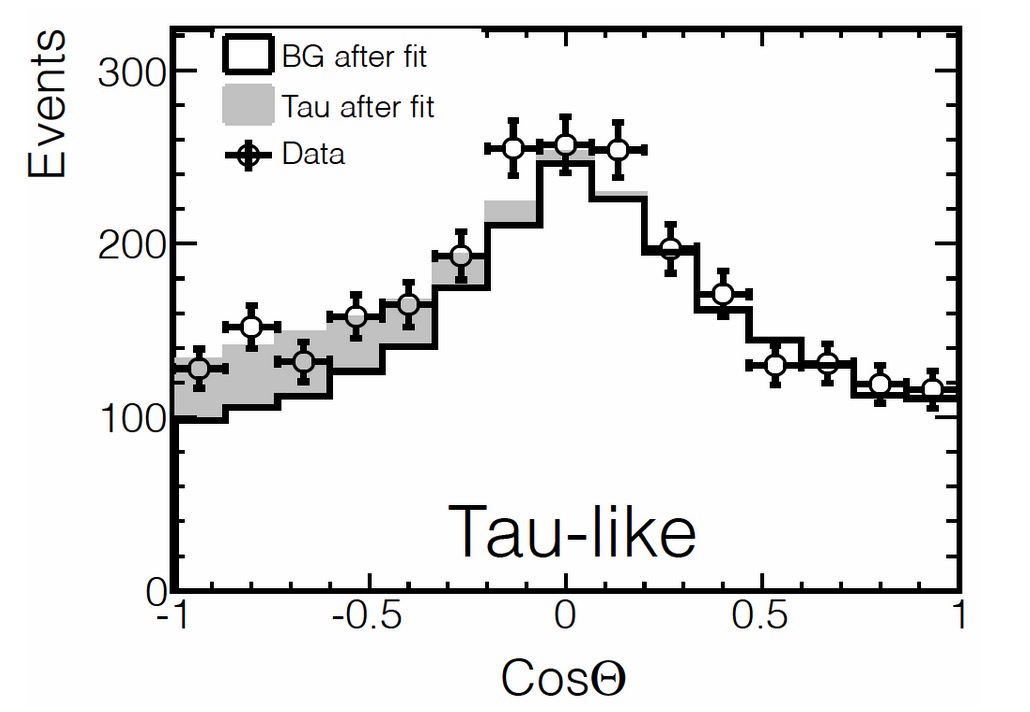}
    \caption{Tau neutrino excess in upward going events in Super-K~\cite{HyperKamiokande:2018ofw}.}
    \label{fig:tau_neutrinos}
\end{figure}

Super-K has demonstrated the ability to identify $\nu_\tau$ in the atmospheric neutrino data, excluding the hypothesis of no-tau-appearance with a significance of 4.6 $\sigma$ (Fig. ~\ref{fig:tau_neutrinos}). After 10 years of running HK, will have the order of 1000~$\nu_\tau$ events that can be used to study CC $\nu_\tau$ cross section, leptonic universality, etc.

With a significant increase of fiducial volume from Super-K, Hyper-K expects to observe 1,000 atmospheric neutrinos in the TeV range. These neutrinos can fill the gap between accelerator-based neutrino experiments and neutrino telescopes, and they open up new opportunities for physics beyond current scopes~\cite{Schneider:2021wzs}. Topics of TeV range atmospheric neutrinos at Hyper-K include neutrino cross-section measurements in the TeVs energy range competitive to the CERN FASER$\nu$ experiment~\cite{FASER:2021mtu}, and provide significant chance to discover new physics including but not limited to TeV boosted dark matter.

In a global context Hyper-K data combined with those from other long-baseline neutrino oscillation experiments will be essential to test non-standard oscillation scenarios, which can only be revealed by combining data from different experiments. 

{\vskip 0.25in}
\noindent
\section{\bf \large Accelerator Neutrinos - The J-PARC to Hyper-K long baseline experiment}
{\vskip 0.075in}

The focus of the J-PARC to Hyper-K experiment is the measurements of $\Delta m^{2}_{32}$, $\sin^{2}\theta_{23}$, $\sin^{2}\theta_{13}$, and $\delta_{CP}$.

\begin{figure}[htb!]
    \centering
    \includegraphics[width=0.75\columnwidth]{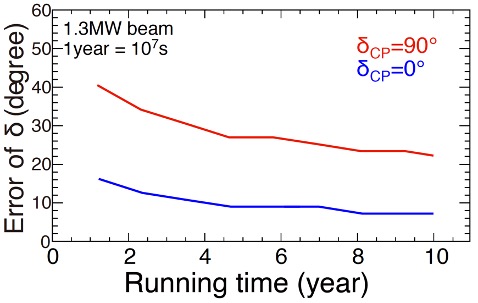}
    \caption{Expected 68\% CL uncertainty of $\delta_{CP}$ as a function of running time. Shown for the normal hierarchy case, with the mass hierarchy assumed to be known~\cite{HyperKamiokande:2018ofw}.}
    \label{fig:Delta_CP_Error}
\end{figure}

The neutrino energy spectrum of the J-PARC neutrino beam is tuned to the first oscillation maximum using the off-axis technique, which enhances the flux at the peak energy while reducing the higher energy component that produces background events. The peak energy, around 600~MeV, is well matched to the water Cherenkov detector technology, which has an excellent e/µ separation capability (miss-identification is less than 1\% for sub-GeV and $\sim$2\% for multi-GeV), high background rejection efficiency and high signal efficiency for sub-GeV neutrino events. Due to the relatively short baseline of 295~km and thus lower neutrino energy at the oscillation maximum, the contribution of the matter effect is smaller for the J-PARC to Hyper-Kamiokande experiment compared to DUNE. Thus, the CP asymmetry measurement with the J-PARC to Hyper-K long baseline experiment has less uncertainty related to the matter effect, while DUNE with its 1,300 km baseline has a much better sensitivity to the mass hierarchy. Nevertheless, as described earlier, Hyper-K can determine the mass hierarchy using atmospheric neutrinos and sensitivities for CP  violation and mass hierarchy can be further enhanced  by combining accelerator and atmospheric neutrino measurements.
Fig.~\ref{fig:Delta_CP_Error} shows the 68\% CL uncertainty of $\delta_{CP}$ as a function of running time (the integrated beam power).

